\documentclass[
 reprint,
amsmath,amssymb,
aps,
longbibliography,
citeautoscript
]{revtex4-1}

\usepackage[Symbol]{upgreek}
\usepackage{amssymb}
\usepackage{mathrsfs}
\usepackage{graphicx}
\usepackage{dcolumn}
\usepackage{bm}
\usepackage{graphicx}
\usepackage{dcolumn}
\usepackage{bm}

\usepackage[english]{babel}
\usepackage{hyperref}
\usepackage{times}
\usepackage{amsfonts,amssymb,bm}
\usepackage{amsmath}
\usepackage{epsfig}
\usepackage{mathrsfs}
\usepackage{color,soul}
\usepackage{bbold}
\usepackage{subfig}
\usepackage{pstricks}
\usepackage{mathtools}
\usepackage{textcomp}
\usepackage{braket}
\usepackage{empheq}
\usepackage{amsmath}
\usepackage{amsthm}
\usepackage[babel]{csquotes}
\usepackage[font=small]{caption}
\usepackage[labelfont=bf,labelsep=quad,justification=raggedright]{caption}
\usepackage{epstopdf}

\begin{document}

\preprint{APS/123-QED}

\title{Classical structured light analogy of quantum squeezed state}

\author{Zhaoyang Wang$^{1,2}$, Ziyu Zhan$^{1,2}$, Xing Fu$^{1,2,\dagger}$, and Yijie Shen$^{3,*}$,}
\affiliation{
	{$^{1}$Key Laboratory of Photonic Control Technology (Tsinghua University), Ministry of Education, Beijing 100084, China}\\
	{$^{2}$State Key Laboratory of Precision Measurement Technology and Instruments, Department of Precision Instrument, Tsinghua University, Beijing 100084, China}\\
	{$^{3}$Optoelectronics Research Centre \& Centre for Photonic Metamaterials , University of Southampton, Southampton SO17 1BJ, UK}\\
}

\date{\today}
\begin{abstract}
Much of the richness in nature arises due to the connection between classical and quantum mechanics. In advanced science, the tools of quantum mechanics was not only applied in microscopic description but also found its efficacy in classical phenomena, broadening the fundamental scientific frontier. A pioneering inspiration is substituting Fock state with structured spatial modes to reconstruct a novel Hilbert space. Based on this idea, here we propose the classical analogy of squeezed coherent state for the first time, deriving classical wave-packets by applying squeezed and displacement operators on free space structured modes. Such a generalized structured light not only creates new degrees of freedom into structured light, including tunable squeezed degree and displacement degree but also exhibits direct correlation between quadrature operator space and real space. Versatile generalized classical squeezed states could be experimentally generated by a simple large-aperture off-axis-pumped solid-state laser. On account of its tunablity, we initially put forward a blueprint using classical structured light, an analogy of squeezed states to realize super-resolution imaging, providing an alternative way to beat diffraction limit as well as opening an original page for subsequent applications of high-dimensional structured light, such as high-sensitive measurement and ultra-precise optical manipulation.     
\end{abstract}

\maketitle

\section{Introduction}
Quantum mechanics was emerged as an increasingly attractive scientific branch due to its so counterintuitive effects in contrast to the classical phenomena. However, in recent advances of structured light~\cite{forbes2020structured}, many quantum optics effects could be also realized in classical intense light~\cite{carusotto2013quantum,konrad2019quantum,cirac2012goals,qian2015shifting,forbes2019classically}. For instance, a classical analogy of Shr\"odinger's cat state can be realized by shaping vortex beams carrying orbital angular momentum (OAM)~\cite{liu2019classical}. The vortex beam can also be used to simulate Landau levels and Laughlin states that reveal topological order in quantum matters~\cite{schine2016synthetic,clark2020observation,corman2020light}. The Berezinskii–Kosterlitz–Thouless phase transition effect in condensed matter can also be simulated by multi-vortex light field~\cite{situ2020dynamics}. In addition to the scalar structured light, the vector beam with spatially non-separable polarization can resemble the quantum entangled Bell states~\cite{kagalwala2013bell,mclaren2015measuring,karimi2015classical,shen2019optical,aiello2015quantum,toninelli2019concepts}. Recently, a novel complex laser modes, exploited to mimic the quantum SU(2) coherent state have been proposed~\cite{shen2020structured,tuan2018characterization,chen2004wave,shen2018periodic,shen2018truncated,shen2018polygonal}. The 
vectorial light state of which can also mimic properties of high-dimensional entangled Greenberger–Horne–Zeilinger (GHZ) states~\cite{shen2021creation}. These implementations of quantum mechanics in classical optics have motivated the development of novel applications in optical (tele)communications~\cite{guzman2016demonstration,ndagano2017characterizing,ndagano2017creation}, optical computing~\cite{goyal2013implementing,goyal2015implementation,sephton2019versatile,d2020two}, 
functional metamaterials~\cite{devlin2017arbitrary,stav2018quantum,wang2018quantum,li2020metalens}, precise metrology and sensing~\cite{d2013photonic,toppel2014classical,berg2015classically}. In short, the exploration of classical correspondence of more quantum states, e.g. cat state, Bell state, and GHZ state, is not only enabling the potential for exciting novel applications, but is also a fundamental scientific endeavour in itself, blurring quantum-classical gap and refreshing human's worldview.

In quantum mechanics, squeezed state is an important quantum state that reveals more generalized distribution of minimum-uncertainty probability wavepacket than the normal coherent state~\cite{walls1983squeezed}, which has been widely used in high-sensitivity interferometry and gravitational wave detection~\cite{aasi2013enhanced,schnabel2017squeezed}. By adding squeezed vacuum state, the temporal redistribution of quantum shot noise has been detected experimentally~\cite{breit2019measure}. Accompanied with time-dependent measurement mechanism as well as lock-phase apparatus, the standard quantum limit (SQL) is suppressed in this way. Though the squeezed vacuum state displays such exciting property,  pure and stable squeezed state is hard to generate. In 2013, squeezed vacuum state was first to be utilized in gravitational wave detection. 
Though squeezed states have made tremendous success in quantum field, to date, the classical structured light analogy of squeezed state was still unexplored yet. Despite classical analogy of complex coherent state which has already been exploited recently to extended multiple degrees of freedom (DoFs) in SU(2) geometric vector vortex beam (VVB)~\cite{shen2020structured}, 
fundamental theoretical framework of such analogy is still lack of generality and complexity. Fortunately, structured modes, such as Hermite-Gaussian (HG) modes, Laguerre-Gaussian (LG) modes exhibit orthogonality and complexity in real space~\cite{beijersbergen1993astigmatic}, corresponding to Fock states in quantum mechanics.  Following the same mathematics as quantum mechanics does, a novel Hilbert space in which varied states represent distinguish classes of VVBs which introduce extra controllable DoFs is established. 
Such ultra-DoF structure in VVB has already holden promises in high-speed laser machining of nano-structures~\cite{omatsu2019new,ni2017three,toyoda2013transfer}, driving the microrobots~\cite{palagi2016structured}, optical tweezers for manipulating particles~\cite{padgett2011tweezers,woerdemann2013advanced,bhebhe2018vector}, high-security encryption~\cite{fang2020orbital}, large-capacity multi-channel communications~\cite{wang2012terabit,bozinovic2013terabit,liu2018direct,willner2018vector}, to name but a few. Moreover, inspired by what squeezed vacuum state plays in gravitational wave detection, classical counterpart might show the same effect in breaking classical Heisenberg limit, namely, diffraction limit. Thus, such structured VVB could be considered as a new approach for super-resolution imaging, providing a powerful tool in astronomy observation, bioscience and material science .

In this paper, we propose the new structured light analogy of coherent squeezed vacuum state, deriving the classical wave-packets by applying squeezed and displacement operators on vacuum state, which provides new tunable DoFs including squeezed degree and displacement degrees. Furthermore, we experimentally generate these modes via an off-axis pumped laser, in good agreement with theoretical simulations. Besides, we handwavingly devise a path-breaking approach to achieve super-resolution imaging based on our theoretical framework, paving the way for further utilization, such as ultra-sensitive measurement, precise optical manipulation, etc.

\section{Theoretical basis: quantum states}
\begin{figure*}
	\centering
	\includegraphics[width=0.8\linewidth]{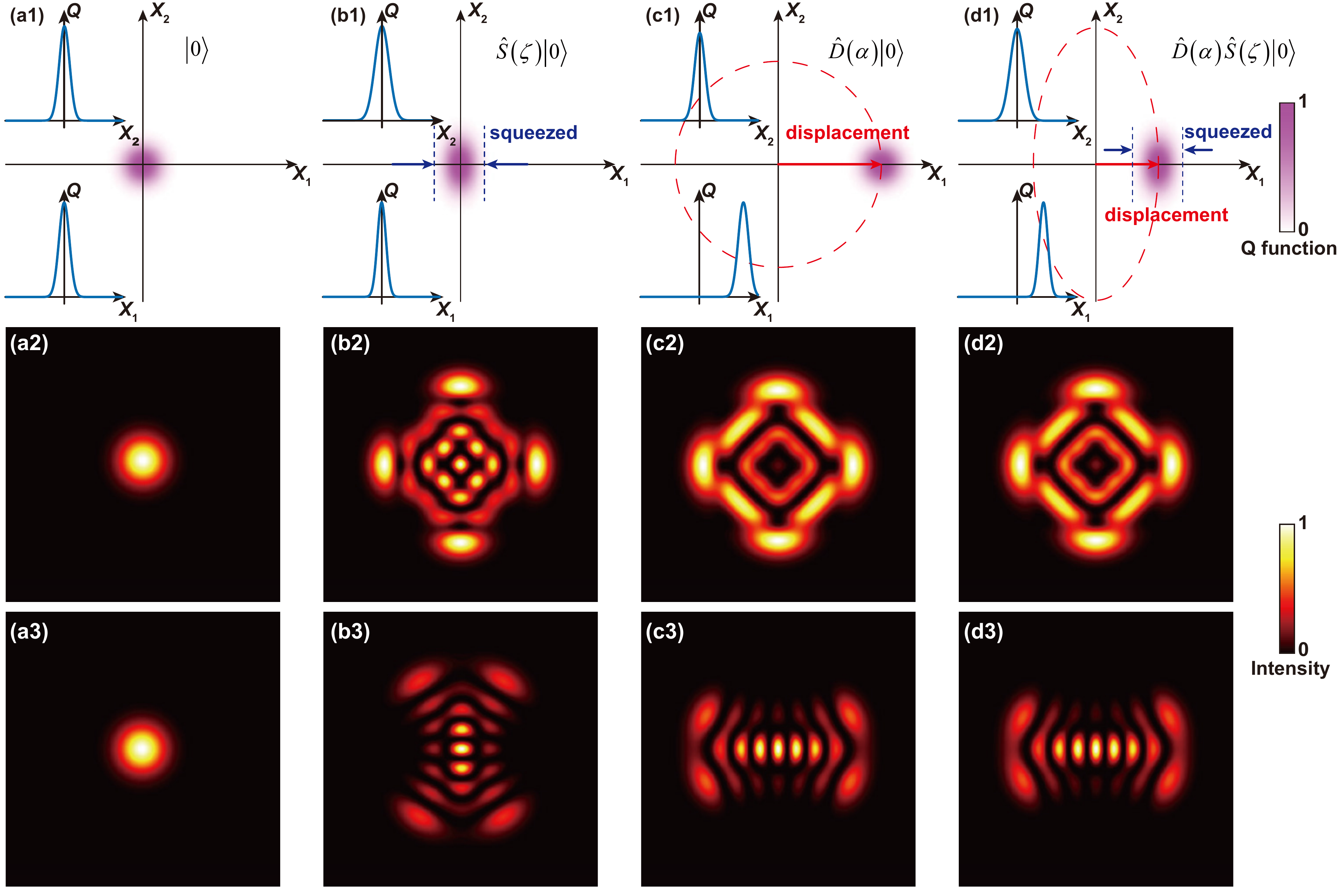}
\caption{$Q$ function and corresponding transverse intensity distributions. Column (a): vacuum state ($N=0$, $r=0$, $\tau=0$); Column (b) SV state ($N=8$, $r=0.5$, $\tau=0$); Column (c) Coherent state ($N=8$, $r=0$, $\tau=6$); Column (d) CSV state ($N=8$, $r=0.5$, $\tau=6$). Subplots on the top row are $Q$ function distributions, where displacement marked with red arrow, squeezing effect marked with blue arrow, red dashed curves in subplots (c1)(c2) represents displacement orbit. Subplots on the middle and bottom rows are transverse intensity distributions based on LG modes and HG modes, respectively.}
\label{f.qfunc}
\end{figure*}

Firstly, several basic definitions such as vacuum state, squeezed vacuum (SV) state, coherent state and coherent squeezed vacuum (CSV) state needs to be clarified. Vacuum state is the minimum uncertainty state with zero average intensity, noted as $|0 \rangle$ commonly. Generally, SV state is obtained by applying squeezed operator on vacuum state as~\cite{2004Introductory,PhysRevD.32.400}:
\begin{align}
	| \zeta \rangle=\hat{S}(\zeta) |0 \rangle,
	\label{s_van}
\end{align}
where $\hat{S}(\zeta)=\exp({\frac{1}{2} \zeta \hat{a}^{\dag 2} -\frac{1}{2} \zeta^{\ast} \hat{a}^{2}})$ is squeezed operator, $\zeta=r \exp({\text{i} \phi})$, $0 \leq r \leq \infty$, $0 \leq \phi \leq 2 \pi$, $r$ determines the squeezing effect when $r \to 0$ no squeezing effect, $\phi$ determines squeezing orientation, $\hat{a}^{\dag}$ and $\hat{a}$ are quantum ladder operators. Consider two quadrature operators defined as $\hat{X}_1=\frac{1}{\sqrt{2}} (\hat{a}^{\dag} + \hat{a})$ and $\hat{X}_2=\frac{\text{i}}{\sqrt{2}} (\hat{a}^{\dag} - \hat{a})$, the uncertainty of their expectations in SV state are expressed as $\Delta^{2}(X_1)=\frac{1}{2}+\sinh^{2}(r) +\sinh(r) \cosh(r) \cos(\phi)$ and $\Delta^{2}(X_2)=\frac{1}{2}+\sinh^{2}(r) -\sinh(r) \cosh(r) \cos(\phi)$, which are not equal in common except limiting case $r \to 0$~\cite{2019New}. The effect of $\hat{S}(\zeta)$ is squeezing the phase-space probability distribution ($Q$ function~\cite{2004Introductory}) in a quadrature direction as shown in Fig.~\ref{f.qfunc} (b1), while the phase-space probability distributions of vacuum state in all directions are equal as shown in Fig.~\ref{f.qfunc} (a1).

By applying displacement operator $\hat{D}(\alpha)$ on vacuum state~\cite{2004Introductory,PhysRevD.32.400}, coherent state can be expressed as:
\begin{align}
	| \alpha \rangle=\hat{D}(\alpha) |0 \rangle,
	\label{d_van}
\end{align}
where $\hat{D}(\alpha)=\exp( {\alpha \hat{a}^{\dag} -\alpha^{\ast} \hat{a}})$, $\alpha=\tau \exp(\text{i}\theta)$, $0 \leq \tau \leq \infty$, $0 \leq \theta \leq 2 \pi$, $\tau$ determines the displacement effect when $\tau \to 0$ no displacement, $\theta$ determines the displacement orientation. The phase-space probability distribution of coherent state on two quadrature components$X_1$ and $X_2$ are equal. The effect of $\hat{D}(\alpha)$ is a displacement (marked with red arrow) and red dashed curves represents displacement orbits, the trajectory of $Q$ function central point with $\tau$ fixed and $\theta$ changed from $0$ to $2\pi$, as shown in Fig.~\ref{f.qfunc} (c1). 

As for CSV state, it can be expressed as~\cite{2004Introductory,PhysRevD.32.400}:
\begin{align}
	| \alpha,\zeta \rangle=\hat{D}(\alpha) \hat{S}(\zeta) |0 \rangle,
	\label{ds_van}
\end{align}
which could also be called displacement squeezed vacuum state due to the effects of squeezing and displacement as shown in Fig.~\ref{f.qfunc} (d1). The phase-space probability distributions mentioned, are represented by the $Q$ function (also called Husimi function), which could provide a intuitive tool to reveal the quantum characters of these states. The $Q$ function of CSV state can be expressed as~\cite{2004Introductory}:
\begin{align}
 & Q(\beta)= \frac{1}{\pi \cosh(r)} \exp \{ -(|\alpha|^2 + |\beta|^2)+\frac{\beta^{\ast} \alpha + \beta \alpha^{\ast}}{\cosh(r)}   \notag \\
 &-\frac{1}{2} [\exp(\text{i}\phi)(\beta^{\ast 2 }-\alpha^{\ast 2})+\exp(-\text{i}\phi)(\beta^{2}-\alpha^{2})]\tanh(r)  \},
\label{q_func}
\end{align}
where $\beta=x+\text{i}y$. We plot $Q$ function in Fig.~\ref{f.qfunc} as (a1) $r=0$, $\tau=0$ for vacuum state; (b1) $r=0.5$, $\tau=0$ for SV state; (c1) $r=0$, $\tau=8$ for coherent state; (d1) $r=0.5$, $\tau=8$ for CSV state. In other words, vacuum state, SV state and coherent state could be seen as three limiting cases of CSV state.

\begin{figure*}
	\centering
	\includegraphics[width=0.8\linewidth]{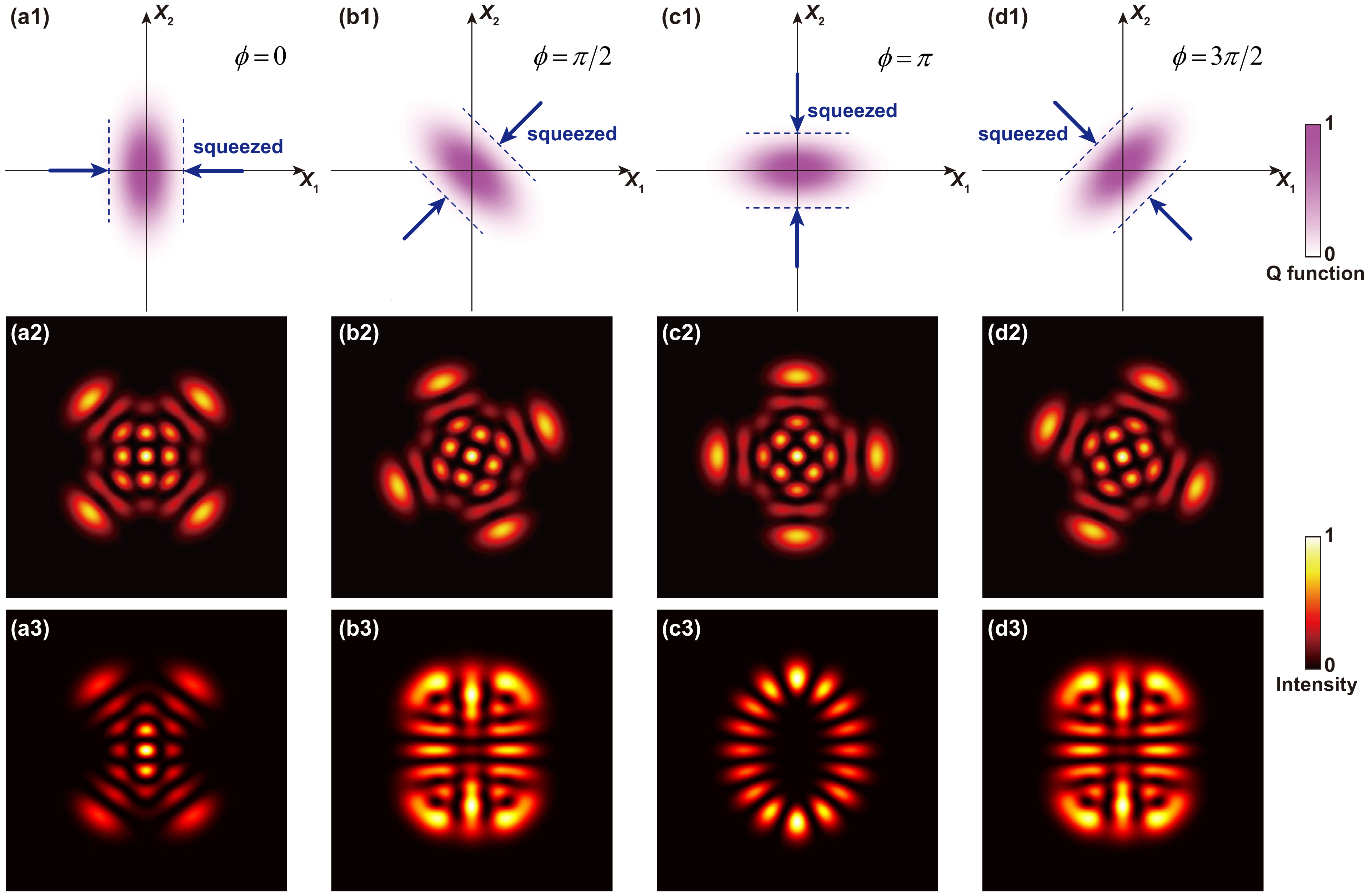}
\caption{The $Q$ function (top row) and corresponding transverse intensity patterns (middle and bottom rows) for SV state. The transverse intensity patterns in middle and bottom rows correspond to classical squeezed light based on LG modes and HG modes, respectively. Squeezing orientations marked with blue arrows and changed with parameter $\phi/2$. Select $N=8$, $r=0.7$ and $\phi=0$, $\pi/2$, $\pi$, $3\pi/2$ from left to right, respectively.}
\label{f.swave}
\end{figure*}

\begin{figure*}
	\centering
	\includegraphics[width=0.8\linewidth]{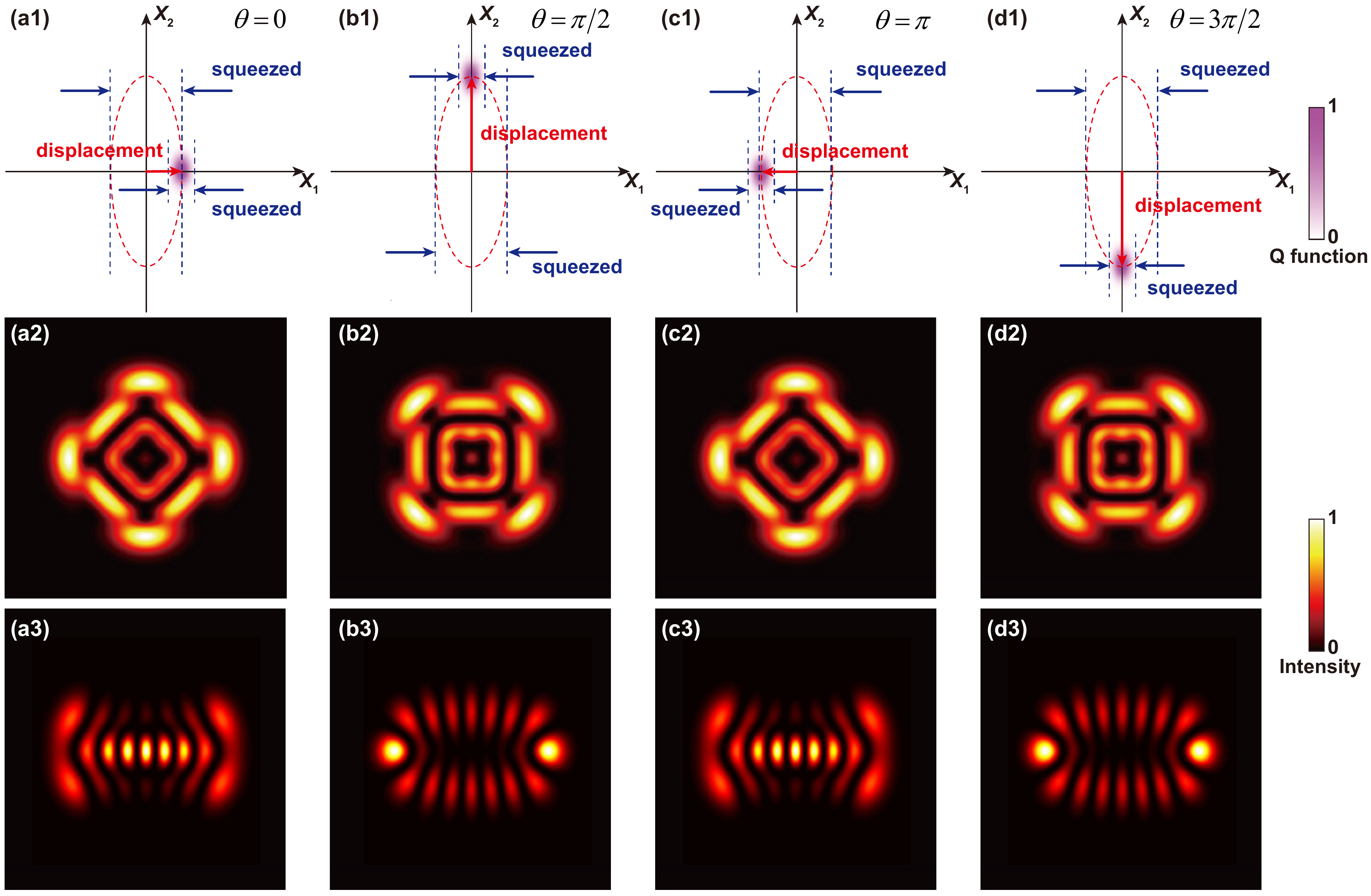}
\caption{The $Q$ function (top row) and corresponding transverse intensity patterns (middle and bottom rows) for CSV state. The transverse intensity patterns in middle and bottom row correspond to classical squeezed light based on LG modes and HG modes, respectively. Displacement orientations marked with red arrows and changed with parameter $\theta$. Squeezing orientations marked with blue arrows and displacement orbits marked with red dashed curves, revealing squeezing effect both in phase-space probability distributions and displacement orbits. Select $N=8$, $r=0.5$, $\tau=6$, $\phi=0$ and $\theta=0$, $\pi/2$, $\pi$, $3\pi/2$ from left to right, respectively.}
\label{f.cswave}
\end{figure*}

\section{Classical wave-packets analogy of quantum states}
Researching the classical structured light analogy of quantum states is an attracting topic, such as classically entangled light~\cite{forbes2019classically,aiello2015quantum}, classical vortex light analogy of quantum cat state~\cite{liu2019classical}, landau levels for photons~\cite{schine2016synthetic} and laughlin states made of light~\cite{clark2020observation}, etc.
In this section, we will derive the classical wave-packet corresponding to quantum states discussed above. Hereinafter, the symbol $|* \rangle$ refers to spatital modes in free space such as vacuum $|0 \rangle$ corresponding to ground Gaussian beam. We focus on the generation of classical wave-packets corresponding to SV state and CSV state. Based on Baker-Campbell-Hausdorff (BCH) relation~\cite{Michael1997Displaced}, squeezed operator $\hat{S}(\zeta)$ could be written as:
\begin{align}
	\hat{S}(s)=\exp\left(\frac{1}{2} s \hat{a}^{\dag 2} \right) (\cosh{(r)})^{-1/2-\hat{a}^{\dag} \hat{a}} \exp \left(-\frac{1}{2} s^{\ast} \hat{a}^{2} \right),
	\label{s_bch}
\end{align}
where $s=\tanh(r)\exp(\text{i}\phi)$. Further considering $\hat{a}^{\dag} \hat{a}=N$ and $\hat{a} |0 \rangle=0$, Eq.~\ref{s_bch} can be derived as:
\begin{align}
	\hat{S}(s) |0 \rangle=[1-\tanh^{2}(r)]^{1/4+N/2} \exp \left(\frac{1}{2} s \hat{a}^{\dag 2} \right) |0 \rangle,
	\label{s_bch2}
\end{align}
Then we could expand the term $\exp(\frac{1}{2} s \hat{a}^{\dag 2})$ by Taylor expansion and derive as:
\begin{align}
	& \hat{S}(\zeta) |0 \rangle = [1-\tanh^{2}(r)]^{1/4+N/2} \notag \\
	& \sum\limits_{M=-J}^{J}{{{\left( \begin{matrix}
			2J  \\
			J+M  \\
			\end{matrix} \right)}^{{1}/{2}}} \left( \frac{s}{2} \right) ^{\frac{J+M}{2}} \frac{\sqrt{(J+M)!}} {\left(\frac{J+M}{2}\right)!} } |J,M \rangle,
	\label{s_bch3}
\end{align}
By substituting $|J,M \rangle$ with eigenmodes such as Hermite-Gaussian (HG) modes or Laguerre-Gaussian (LG) mode, we could derive the classical wave-packets corresponding to SV state as:
\begin{align}
	& \Psi^{\text{SV}}(x,y,z|r,\phi) = [1-\tanh^{2}(r)]^{1/4+N/2} \notag \\
	& \sum\limits_{K=0}^{N}{{{\left( \begin{matrix}
			N  \\
			2K  \\
			\end{matrix} \right)}^{{1}/{2}}} \frac{\tanh^{K}(r)}{2^{K}} \text{e}^{\text{i}K\phi} \frac{\sqrt{(2K)!}} {K!} } \psi^{\text{eigen}}_{N-2K,2K}(x,y,z),
	\label{s_wave}
\end{align}
where $ \psi^{\text{eigen}}_{n,m}(x,y,z)$ represents eigenmodes with indices $(n,m)$. 
The analogy of quantum inner product $\langle n,m|\zeta \rangle$ could be expressed as $\psi^{\text{eigen} \ast}_{n,m} \Psi^{\text{SV}}$, i.e. the probability of sub-state $|n,m \rangle$ in SV state $|\zeta \rangle$ corresponding to the weight of eigenmode $\psi^{\text{eigen}}_{n,m}$ in wave-packet $\Psi^{\text{SV}}$.
The classical wave-packets of CSV state can be derived analogously as:
\begin{align}
	\hat{D}(\alpha)\hat{S}(\zeta) |0 \rangle =& [1-\tanh^{2}(r)]^{1/4+N/2} (1+\tau^2)^{-N/2} \notag \\
	& \exp( \alpha \hat{a}^{\dag}) \exp\left(\frac{1}{2} s \hat{a}^{\dag 2}\right) |0 \rangle,
	\label{s_bch4}
\end{align}
By exploiting Taylor expansion and substituting $|J,M \rangle$ with eigenmode $\psi^{\text{eigen}}$, the classical wave-packets of coherent squeezed state can be derived as:
\begin{align}
	& \Psi^{\text{CSV}}(x,y,z|r,\phi,\tau,\theta) = [1-\tanh^{2}(r)]^{1/4+N/2} (1+\tau^2)^{-N/2} \notag \\
	& \sum\limits_{K_1=0}^{N} \sum\limits_{K_2=0}^{K_1}{{{\left( \begin{matrix}
			N  \\
			K_1  \\
			\end{matrix} \right)}^{{1}/{2}}} \left[ \frac{\tanh(r) \text{e}^{\text{i} \phi}}{2}\right] ^{\frac{K_1 - K_2}{2}} \frac{\sqrt{K_1!} \tau^{K_2} \text{e}^{\text{i}K_2 \theta}} {K_2! \left(\frac{K_1 -K_2}{2}\right)!} } \notag \\
		&	\psi^{\text{eigen}}_{N-K_1,K_1}(x,y,z),
	\label{cs_wave}
\end{align}
where the weight of eigenmode $\psi^{\text{eigen}}_{n,m}$ in wave-packets $\Psi^{\text{CSV}}$, $\psi^{\text{eigen} \ast}_{n,m} \Psi^{\text{CSV}}$, corresponds to quantum inner product $\langle n,m|\alpha, \zeta \rangle$. The transverse distributions of Eq.~(\ref{cs_wave}) in real space are analogous to $Q$ function distribution in phase space, as shown in Fig.~\ref{f.qfunc}, where (a2)(a3) select $r=0$, $\tau=0$, $N=0$ for classical analogy of vacuum state; (b2)(b3) $r=0.5$, $\tau=0$, $N=8$ for classical analogy of SV state; (c2)(c3) $r=0$, $\tau=8$, $N=8$ for classical analogy of coherent state; (d2)(d3) $r=0.5$, $\tau=8$, $N=8$ for classical analogy of CSV state. Analogously, the classical analogies of vacuum state, SV state and coherent state could be seen as three limiting cases of classical analogy of CSV state. Transverse intensity distributions on middle and bottom rows are calculated by selecting $\psi^{\text{eigen}}_{n,m}(x,y,z)$ as $\text{LG}_{\rho,\ell}(x,y,z)$ ($\rho=\min(n,m)$, $\ell=n-m$) and $\text{HG}_{n,m}(x,y,z)$, respectively.

More detailing analogies of SV state and CSV state are shown in Fig.~\ref{f.swave} and Fig.~\ref{f.cswave}, respectively. The parameter $\phi$ determines squeezing orientation as shown in top row of Fig.~\ref{f.swave}. Transverse intensity distributions based on LG modes corresponds spatial rotation determined by parameter $\phi$ as shown in middle row of Fig.~\ref{f.swave}. Transverse intensity distributions based on HG modes corresponds spatial pattern evolution related to parameter $\phi$ as shown in bottom row of Fig.~\ref{f.swave}. The squeezing effects in CSV state are reflected on shapes both of $Q$ function and displacement orbits and displacement orientation are determined by parameter $\theta$, as shown in top row of Fig.~\ref{f.cswave}. Transverse intensity distributions based on LG modes corresponds spatial rotation determined by parameter $\theta$ as shown in middle row of Fig.~\ref{f.cswave}. Transverse intensity distributions based on HG modes corresponds spatial pattern evolution related to parameter $\theta$ as shown in bottom row of Fig.~\ref{f.cswave}.

\section{Experiments of generating classical squeezed state light}
\begin{figure}
	\centering
	\includegraphics[width=\linewidth]{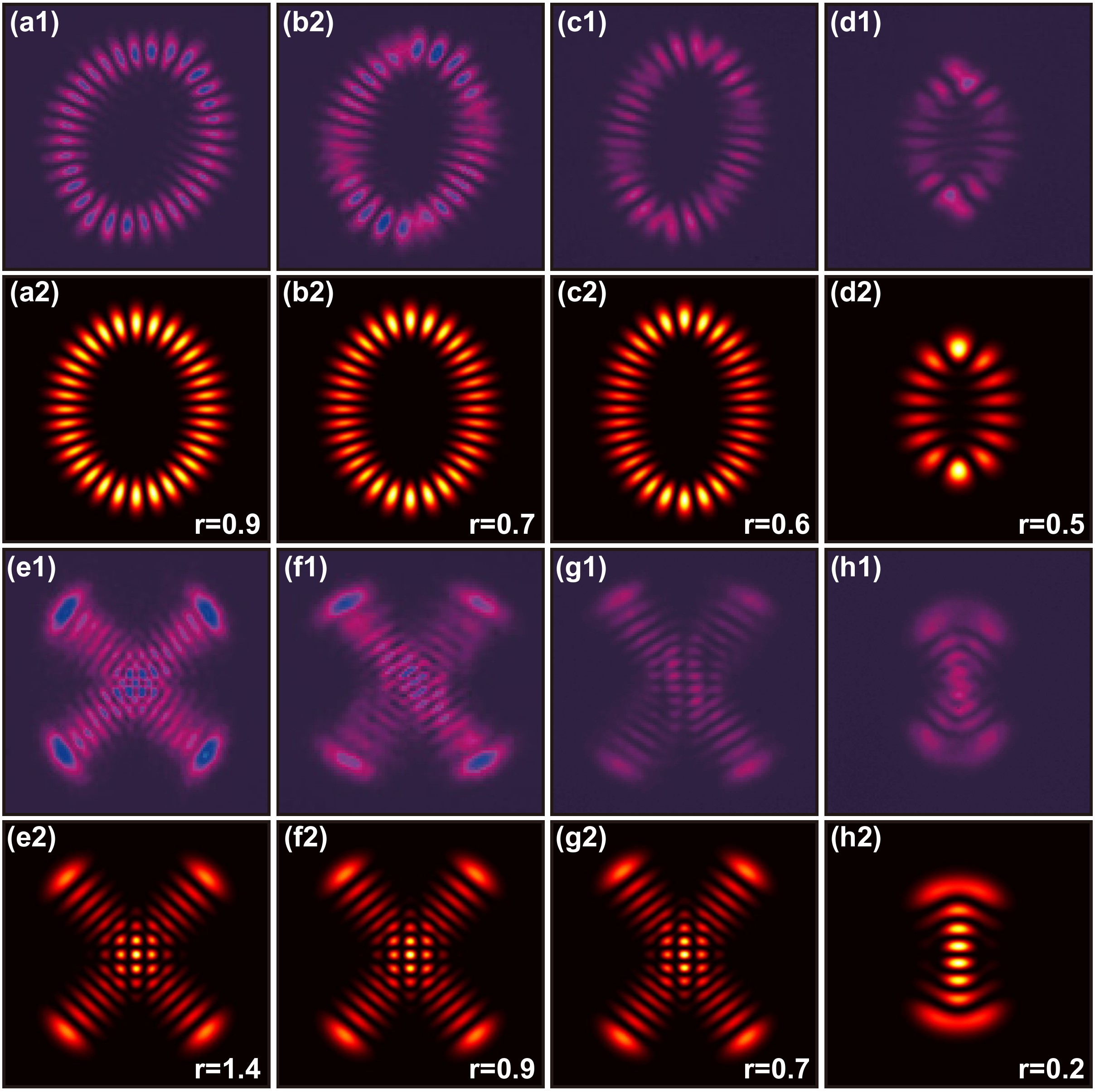}
\caption{Experimental results of generating classical squeezed state light (first and third rows) and corresponding simulations based on HG modes (second and forth rows). The simulating parameters from left to right are $N=$(16, 16, 16, 7, 16, 16, 16, 7), $r=$(0.9, 0.7, 0.6, 0.5, 1.4, 0.9, 0.7, 0.2), $\phi=\pi$ (a--d) and $0$ (e--h), respectively.}
\label{f.expsim}
\end{figure}

We generate classical SV state light via a large-aperture off-axis-pumped solid-state laser. Our theoretical model can explain the laser emitting mechanics with good agreement. Some results and corresponding simulations based on Eq.~\ref{s_wave} are shown in Fig.~\ref{f.expsim}. The experimental setup included a 976 nm fiber-coupled laser diode (LD) (Han’s TCS, core: 105 um, NA: 0.22) as the pump source, two identical convex lens (focal length is 60 mm) as the pump coupling system (magnification is about 2:1), a 4$\times$4$\times$2 mm$^3$ a-cut thin-slice Yb:CALGO (Altechna, 5 at.$\%$) as the gain medium, a flat dichroic mirror, antireflective (AR) coated at 976 nm, high-reflective (HR) coated at 1020–1080 nm, and a concave output mirror (OC: transmittance of 1$\%$ at 1020–1080 nm, the radius of curvature $R=$100 mm) as the cavity mirrors. The crystal was embedded in a copper heat sink water-cooled. CCD was used to record experimental results.

\section{Discussions}
In this work, we propose a generalized mathematical framework akin to that of quantum mechanics, in which Fock states are replaced by structured modes in free space. 
Afterwards, we put forward the classical counterpart of quantum squeezed states whose pattern of $Q$ function in phase diagram shows strong correlation with spatial intensity pattern in real space. 
Based on the intuition of applying quantum squeezed state in precise measurement, here we design a novel spatial super-resolution imaging approach based on classical structured light.
 Before further discussion, it is necessary to make sure the key-points for precise detecting system applied SV state. As for ultra-precise measurement via interferometer for frequency shift, displacement, its sensitivity is ultimately constrained by SQL causing from quantum shot noise. Such limitation could straightforwardly be suppressed by enhancing detection source intensity. Nevertheless, increasing source power inevitably introduces additional disturbation resulting from nonlinear effects, more violent mechanics vibration, etc. An alternative approach to promote the resolution of system is injecting squeezed vacuum state instead of classical vacuum state. In turn, quantum shot noise exhibits temporal periodicity. Hence, with the help of locking phase apparatus, if the detection time at which the quantum noise is ``squeezed'' is elaborately set, the SQL of detection will be suppressed. 
 
Akin to the correspondent relation between classical and quantum entanglement, here we follow a resemble path as the method of detecting for gravitational wave using quantum squeezed state to establish classical measurement system. In our mathematical framework, the pattern of $Q$ function which reveals quantum characteristics in quadrature operator space coincides with intensity pattern of structured modes in real space. For instance, $Q$ function for vacuum state is Gaussian shape which perfectly aligns with our classical vacuum light. Further, the pattern for number state exhibits ring-like shape, corresponding to the intensity pattern for classical number state formed by LG modes. Therefore, the uncertainty for quadrature operator could be directly mapped onto spatial uncertainty of intensity, or in other words, spatial resolution limited by diffraction limit. Explicitly, in certain orientation, spatial resolution beats the diffraction limitation, while in the orthogonal orientation spatial pattern seems more ``blurred'' than before. It is reasonable to merely extract information in certain orientation which exhibits highest spatial resolution while neglecting others. By applying  detecting beams with different preference of spatial variance, it is possible to retrieve entire spatial information for certain spatial shape.




\bibliography{apssamp}
%

\end{document}